 \documentstyle[12pt]{article}
\begin{document}
\newcommand{\bea}{\begin{eqnarray}}
\newcommand{\eea}{\end{eqnarray}}
\newcommand{\be}{\begin{equation}}
\newcommand{\ee}{\end{equation}}
\newcommand{\gf}{\gamma_{5}}
\newcommand{\gm}{\gamma_{\mu}}
\newcommand{\gn}{\gamma_{\nu}}
\newcommand{\x}{\psi}
\newcommand{\xx}{\bar{\psi}}
\newcommand{\mesx}{\prod d \psi d \bar{\psi}}
\newcommand{\mesa}{\prod_{\kappa} d a_{\kappa} d a_{\kappa}^{\ast}}
\newcommand{\fmn}{F_{\mu \nu}}
\newcommand{\ph}{\phi_{\kappa}}
\newcommand{\phb}{\bar{\phi}_{\kappa}}
\newcommand{\phbp}{\bar{\phi}_{\kappa'}}
\newcommand{\php}{\phi_{\kappa'}}
\newcommand{\lk}{\lambda_{\kappa}}
\newcommand{\tra}{tr(t_{\delta}\{t_{\alpha},t_{\beta}\})}
\newcommand{\tta}{T_{\alpha}}
\newcommand{\ttb}{T_{\beta}}
\newcommand{\axi}{\frac{(1+\gf)}{2}}
\newcommand{\gr}{\gamma_{\rho}}
\newcommand{\gs}{\gamma_{\sigma}}
\newcommand{\dsl}{D \hspace{-0.10in}/}
\newcommand{\dlt}{\delta^{4}}
\newcommand{\nk}{n_{\kappa}}
\newcommand{\ins}{\int d^{4}x}
\newcommand{\inm}{\int \frac{d^{4}k}{(2\pi)^{4}}}
\newcommand{\lag}{{\cal L}}
\newcommand{\anom}{{\cal F}}
\newcommand{\dir}{{\cal M}_\psi}
\newcommand{\ta}{t_{\alpha}}
\newcommand{\tb}{t_{\beta}}
\newcommand{\trab}{tr(t t_{\alpha} t_{\beta})}
\newcommand{\trabd}{tr(T_{\delta} \{T_{\alpha}, T_{\beta} \})}
\newcommand{\ffdm}{F_{\mu \nu}^{\alpha} F_{\rho \sigma}^{\beta}
            \epsilon^{\mu \nu \rho \sigma}}
\newcommand{\ffd}{tr(F_{\mu \nu} \; ^{\ast}F_{\mu \nu})}
\newcommand{\ak}{a_{\kappa}}
\newcommand{\aks}{a_{\kappa}^{\ast}}
\newcommand{\pasl}{\partial \hspace{-0.085in}/}
\newcommand{\sbd}{\bar{s}_\mu \partial_\mu}
\newcommand{\et}{\eta^a _{\mu \nu}}
\newcommand{\radi}{(r^2 + \rho^2 _I)}
\newcommand{\radh}{(r^2 + \rho_H ^2)}
\newcommand{\xin}{(x^2 + \rho_I ^2)}
\newcommand{\xh}{(x^2 + \rho_H ^2)}
\newcommand{\asl}{A^3 \hspace{-0.1in}/}
\newcommand{\wpsl}{W^+ \hspace{-0.1in}/}
\newcommand{\wmsl}{W^- \hspace{-0.1in}/}
\newcommand{\sd}{s_\mu \partial_\mu}
\newcommand{\sx}{s_\nu x_\nu}
\newcommand{\sbx}{\bar{s}_\nu x_\nu}
\newcommand{\jf}{J_{\mu}^{5}}
\newcommand{\ri}{\rho_I}
\newcommand{\rh}{\rho_H}
\newcommand{\ath}{\tanh^{-1}z}
\newcommand{\mw}{M_W ^2}

\vspace{0.8in}
\title{On Fermion Zero-Modes in Instanton $V-A$ Models With
Spontanous Symmetry Breaking}
\author{Kamran Saririan\thanks{E-mail: saririan@utaphy (bitnet), or
saririan@utaphy.ph.utexas.edu}}
\date{}
\maketitle

\centerline{Department of Physics}
\centerline{University of Texas}
\centerline{Austin, Texas 78712}
\vspace{1in}

\begin{abstract}
The left and right handed fermion
zero-modes are examined. Their behaviour
 under the variation of the size of the instanton, $\ri$,
 and the size of the Higgs core,
 $\rh$, for a range of Yukawa couplings corresponding to the fermion
masses in the electroweak theory are studied.
 It is shown that the characteristic radii of the zero-modes, in particlar
those the left handed fermions,
are locked to the instanton size, and are not affected by the variation of
$\rh$, except for fermion masses much larger than those  in the standard
electroweak theory.
\end{abstract}
\newpage
It was realised in the early studies of instanton physics, that zero modes of
fermions play an essential role in theories with topological configurations of
gauge fields \cite{th,th2,cdg,jr} (for a review on instantons, see ref.
\cite{sc}).
For instance, the amplitude of tunnelling between vacua $|n>$ and $|n+1>$
vanishes in presence of any fermion zero modes.
 The existence of fermion zero
modes, also gives rise to fermion number
 violation in the instanton-induced vertices.
This can be seen by considering the matrix element of a product of bilinears of
Fermi fields in an instanton background:
\[ <\anom(\x)>_{\nu=1} = <\xx_1\x_1 \cdots \xx_j\x_j>_{\nu=1}
 \] (in general, each factor may be of the form $\xx_iJ_i\x_i$). If the
Euclidean fermion action is \(I_F= \int d^4x \sum_j \xx_j {\cal M}_{\x_j}
 \x_j \) with
\(j=1,\dots,N_F\), then
\bea <\anom> & \propto & \int _{\nu=1} {\cal D}A \int \mesx \anom(\x)
e^{-I_{YM}
- I_F} \nonumber \\  & \propto & \int_{\nu=1} {\cal D} A \int \mesa \anom(\x)
 e^{-\sum_{\kappa} \lk \ak
\aks} e^{-I_{YM}}, \nonumber \eea where we have expanded in the
modes $\ph$ of $\dir$, given by \(\dir \ph = \lk \ph\),
and \( \x = \sum \ak \ph \) (the flavour indeces have
been suppressed). Since the exponential
only contains contributions from \(\lk \neq 0\), we write:
\[
<\anom> \propto \int \prod_{\lk\neq 0} d\ak d\aks \left[
\int \prod_{\lk=0}^{N_0} d\ak^0 d\ak^{0\ast} \anom (a)\right] e^{-\sum_{\lk
\neq 0} \lk \ak \aks} . \]
Indeed, the Grassmannian integrals in the brackets are nonvanishing only
if $\anom(\x)$ is a product of precisely $N_0$ bilinears of different flavours.
In fact for winding number $\nu=1$, by virtue of the index theorem \cite{egh}
and the presence of the anomalous chiral fermion current, the number of the
fermion zero modes is equal to $N_F$. So, in the instanton-induced fermion
 vertices, there is chiral charge violation of $\Delta Q_5 = 2N_F$. This leads
to baryon and lepton number violations (albeit, suppressed by factors of  the
order \(\sim e^{-8\pi^2/g^2}\)). There has been  a considerable
recent interest in the possibility that the baryon number in the universe
was generated by the above mechanism, at the scale of electroweak interactions
(see ref. \cite{mattis} and the references therein).  In such a scenario,
the fermion zero-modes are clearly crucial.

The zero modes of fermions in $V-A$ theories, such as the Standard Model of
Weinberg and Salam have previosly been studied \cite{krt}. In this case,
due to Higgs symmetry breaking, the fermions are massive. And the $V-A$ nature
of the theory makes it nontrivial to calculate, exactly,
 the zero-modes except for
some simple cases (for example, spherically symmetric {\sl ansatz}, and equal
 masses of the up-like and down-like quarks \cite{krt}). Furthermore, it has
been assumed that the size of the instanton and that of the approximate
Higgs solution \cite{aff} are equal. Indeed, there is no {\it a priori} reason
that this should be the case.

In this paper, we shall investigate  the fermion zero modes
  in the Weinberg-Salam model with gauge field configurations of
winding number $\nu=1$, in the case that the Higgs scale, $\rh$, and
the instanton scale, $\ri$, are not equal. We address the following
question: Are the scales of the (left handed and right handed)
zero modes ``locked'' to $\ri$ or $\rh$?
Or alternatively, which scale determines the characteristic size
of the zero-modes? We examine  the
behaviour the left-handed and right-handed  zero mode solutions (the
latter do not couple to the  instanton, but only to the Higgs) for a
wide range of Yukawa coupling, corresponding to fermion masses
$\sim 1MeV$ to $\sim 100GeV$.

Let us start with the Lagrangian for the $SU(2)\times U(1)$ Standard
Model:
\bea
\lag &=& - \frac{1}{4}tr(F_{\mu \nu}F^{\mu \nu})
-\frac{1}{4}G_{\mu\nu}G^{\mu\nu} - \bar{L}\dsl L -\bar{R} \dsl R
\nonumber \\
&+& (D_\mu \Phi)^{\dagger} D^\mu \Phi - V(\Phi) -
g_e\{ (\bar{L} \Phi)R -\bar{R}(\Phi^{\dagger} L)\}
\eea
where \( \Phi=\left( \begin{array}{c} \phi^+ \\ \phi^0 \end{array} \right)
 \) is the Higgs doublet, $L= \left( \begin{array}{c} \nu \\ e_L
\end{array} \right)$ is the fermion doublet, $R=e_R$ is the right handed
singlet\footnote{We choose to work with one lepton
family. This simplifies the algebra, since there would be an additional
 equation for the second right handed fermion, identical in form to that
of the first one.},
and $V(\Phi)$ is the Higgs potential.
The part of the Lagrangian that concerns us is the part that involves
the fermions, excluding their $U(1)$ terms.  The fermion equations of motion
after the symmetry breaking are :
\bea
i \pasl \nu + g \gamma.A \nu  &=& 0  \nonumber
 \\ i \pasl e_L + g \gamma.A e_L &=& g_e \phi^{0 \dagger} e_R \nonumber
 \\  i \pasl e_R - g_e \phi^0 e_L &=& 0.
\eea

We use the chiral representation of the $\gamma$-matrices:
\[ \gamma^\mu =\left( \begin{array}{cc}  0 & \sigma^\mu \\ \bar{\sigma}^\mu
 & 0 \end{array} \right), \hspace{0.4in} \gf= \left( \begin{array}{cc}
-i &  0 \\  0 & i \end{array} \right) \]
where \( \sigma^\mu = (\sigma^0, \vec{\sigma}) \), \( \bar{\sigma}^\mu
 = (\sigma^0,
-\vec{\sigma}) \), \(\sigma^0 = i1\hspace{-0.05in}1 \), and metric: \(
\eta_{\mu \nu} = Diag(-1, +1, +1, +1) \). To Euclideanize, we simply let
\bea
\sigma^\mu & \rightarrow & s_\mu =(s_4, \vec{s}) = (-i, \vec{\sigma})
\nonumber \\
\bar{\sigma}^\mu & \rightarrow & \bar{s_\mu} = (-i, -\vec{\sigma})
 = -s^{\dagger}_\mu .    \setcounter{equation}{3}
\eea
In this representation, we may express the fields in terms of
two component spinors:
\be
\nu_L=\left( \begin{array}{c} \alpha \\ 0 \end{array} \right); \hspace{0.2in}
e_L=\left( \begin{array}{c} \beta \\ 0 \end{array} \right); \hspace{0.2in}
e_R=\left(\begin{array}{c} 0 \\ \epsilon \end{array} \right),
\ee
 and re-write equations (2) as follows:
\bea
i \sbd \psi(x)  + \frac{g}{2} \bar{s}_\mu A_\mu^a \tau^a \psi(x)
&-& g_e \phi^{0 \ast}(x) \chi(x) = 0 \nonumber \\
i \sd \chi(x) &-& g_e \phi^0 (x) \x (x) = 0
\eea
where
\[ \psi_{\alpha s}
= \left( \begin{array}{c} 0 \\ \alpha \\ 0 \\ \beta \end{array} \right)  ;
\hspace{0.2in} \chi_s = \left( \begin{array}{c} 0 \\ \epsilon \end{array}
\right) = e_R .
\]
The index $\alpha$ is the gauge group index, and $s$ is the spinor index.

In Higgs theories, the exact instanton solutions cease to exist,
however, approximate solutions can be found \cite{th2,aff}, and can
be shown to have the same winding number as the gauge field \cite{rf}. These
approximate solutions exist at scales $\rh, \ri \ll 1/v$, where $v$ is
the Higgs vacuum expectation value.
These solutions do not correspond to the minimum of the Yang-Mills plus
Higgs action (although they correspond to saddle points of the action). They
can be regarded as the minimum action solutions for a more fundamental theory
at a higher energy scale than the EW-scale, whose Lagrangian may contain
terms of higher order in derivatives suppressed by an inverse power of a
large mass.\footnote{I wish to
thank S. Rajeev for the discussion on this point.}
In the finite gauge, these
solutions (located at the origin)
are given by\footnote{Here,  $\et$ are the 't Hooft symbols:
$\et = \epsilon_{a \mu \nu}$ for $\mu,\nu=1,2,3$ ;
\( \eta^a _{4 \nu}=-\delta_{a \nu}\); \(\eta^a _{\mu 4} = \delta_{a \mu}\);
and \(\eta^a _{44}=0\).}:
\bea
A_\mu &=& -i\frac{g}{2} \tau^a A^a _\mu = \frac{\tau^a \et x_\nu}{\xin} \\
\phi^0 &=&  \frac{i \bar{s}_\mu x_\mu}{\xh^{1/2}}
\eea
If we insert these into equation (5) and carry out some algebraic steps
using identities such as \(-(\sx).(\sbd)=x_\mu \partial_\mu -
2\vec{L}_1 .\vec{\tau}\), and \( -(\sbx)(\sd)= x_\mu \partial_\mu -
2\vec{L}_2 .\vec{\tau} \), we arrive at\footnote{The definitions of
angular momentum operators $L_1$ and $L_2$ are also those given by 't Hooft
\cite{th2}: \(L_{1a} =-\frac{1}{2} i \et x_\mu \partial_\nu\), and
\(L_{2a}=-\frac{1}{2}i\bar{\eta}^a _{\mu \nu} x_\mu \partial_\nu\).}:
\bea
\frac{1}{x^2}(x_\mu \partial_\mu - 2\vec{L}_1 .\vec{\tau})\x &+&
\frac{3}{\xin} \x = \frac{g_e v}{(\xh)^{1/2}}\chi \nonumber \\
\frac{1}{x^2}(x_\mu \partial_\mu - 2\vec{L}_2 . \vec{\tau})\chi &=&
\frac{g_e v}{\xh^{1/2}}\x  .
\eea
We further assume that $\x$ and $\chi$ are spherically symmetric ({\it
i.e.,} $\vec{L}_1 = \vec{L}_2 = 0$):
\be
\x=p(r)u_{0 \alpha s}; \hspace{0.2in} \chi=q(r)w_{0 s} \ee
where $u_0$ and $w_0$ are constant vectors.
Now the coupled radial equations
 read:
\bea
\frac{dp}{dr^2} &+& \frac{3p}{2 \radi} = \frac{g_e v q}{2 \radh^{1/2}}  \\
\frac{dq}{dr^2} &=& \frac{g_e v p}{2 \radh^{1/2}}
\eea
The second order equation for q is:
\be
\frac{d^2q}{d(r^2)^2} + \left( \frac{3}{2} \frac{1}{\radi} + \frac{1}{2}
\frac{1}{\radh}\right) \frac{dq}{dr^2} - \frac{b^2}{\radh}q =0
\ee
where $b=g_ev/2$. Notice that in the limit $\ri=\rh=\rho$, this equation
 reduces to \(q''+(2/y)q'-(b^2/y)q=0\), where $y=r^2+\rho^2$. This equation
can be solved by standard methods, yielding the following
normalisable\footnote{
There are also solutions that are not normalisable, namely those obtained by
replacing the $K_\lambda$ Bessel functions with $I_\lambda$ Bessel functions
in equations 13 and 14. These are
eliminated by imposing boundary conditions $p,q\rightarrow 0$ as $r \rightarrow
0$.}
solutions \cite{krt}:
\bea
q &=& \frac{-Cg_e v \rho}{(r^2 +\rho^2)^{1/2}} K_1(g_ev(r^2 + \rho^2)^{1/2})
\\
p &=& \frac{Cg_e v \rho}{(r^2 +\rho^2)^{1/2}} K_2(g_ev(r^2 +\rho^2)^{1/2})
\eea
For \(\ri\neq\rh\), we may get some qualitative features of the solutions by
looking at the factorised form of $q$. We let:
\be
 q =h(r^2)f(r^2) \ee
such that
\bea h' &+& {\cal P}h = 0  \nonumber   \\
f''&+& ({\cal Q} -\frac{1}{2}{\cal P}' -
 \frac{1}{4}{\cal P}^2)f=0 \nonumber \\
\eea where ${\cal P}$ and ${\cal Q}$ are given by:
\bea {\cal P} &=& \frac{3}{2}\frac{1}{\radi} + \frac{1}{2}
\frac{1}{\radh} \nonumber
\\  {\cal Q} &=& b^{2}/\radh   ,  \nonumber \eea
and find that
\be h = e^{-\frac{1}{2}\int^{r^2}{\cal P}(\xi)d\xi}
=\frac{1}{\radh^{1/4}\radi^{3/4}} . \ee
Now, $f$ satisfies:
\be f'' + \left[\frac{3}{16}\left(\frac{1}{\radi}+
\frac{1}{\radh}\right)^2 -\frac{b^2}{\radh}\right]
f=0.
\ee
For small Yukawa couplings, $\ri$ and $\rh$ come in equal footing in
the equation for $f$; {\em i.e.,} in this limit,
 $f$ is equally sensitive to
 variations in $\rh$ and $\ri$. Furthermore, as the Yukawa coupling gets
stronger, due to  the contribution of the \(b^2/\radh\) the variations of
the Higgs core are expected to affect the zero-modes more. But
 for physically observed fermion masses in the standard
model this effect is small compared to the effect due
 to the quadratic term in
equation (18). So, $f$ is approximately
 equally sensitive to variations of
$\rh$ and $\ri$. But \(h=\radh^{-1/4}\radi^{-3/4}\) is
 clearly more sensitive
to change in $\ri$.  Thus we expect a stronger correlation between the
characteristic radii of the zero-modes and the
 instanton size than between the
characteristic radii of zero-modes and the Higgs core.  On physical
grounds, the variation of the Higgs core is expected to
affect the right handed zero-modes more (if any), because the
the right handed fermions are coupled to the Higgs and not to
the instanton.

These
behaviours of the zero mode solutions were studied
 numerically. The function
\( z=\tanh(r^2M_{W}^2)\) was used to map the radial
 coordinate $r$ to the
unit interval,  and the boundary conditions
 were imposed near \(z=1\). This
simplifies the numerical computations, and
is advantageous because small values of $r^2$ are mapped almost
linearly, and , indeed, we are
interested in the behaviour of the  solution
 at length scales  of the two
cores which are typically small  due to the
 restriction for existence of
approximate solutions. The corresponding equation
 for $q$ in the
$z$-coordinate is given by:
\bea
\frac{d^2q}{dz^2} & + & \frac{1}{1-z^2} \left[ -2z  +
 \frac{3}{2(\ath +\mw \ri^2)}
+\frac{1}{2(\ath + \mw \rh^2)} \right] \frac{dq}{dz}  \nonumber \\
& - & \frac{b^2}{\mw (1-z^2)^2
(\ath + \mw \rh^2)}q = 0 .
\eea
Fig-1($a$) and 1($b$)  show the solutions of the right
handed and left handed fields
respectively for fixed \(\ri=2\times10^{-4}GeV^{-1}\) and four $\rh$
values in the  range:
  \[ 10^{-2}\ri\leq\rh\leq\ri \]  for various $m_F$.
 Notice that the left handed
zero-mode is unchanged as we decrease $\rh$ over two orders of
magnitude.  For a larger instanton, \(\ri=10^{-3} GeV^{-1}\),
 the solutions are shown in Figures 2 and 3,
with the Higgs core, again, given by \( 10^{-2}\ri\leq\rh\leq\ri \).
The left handed fermion is again,  unaffected, except for the
very heavy fermion of $m=500GeV$ (Fig- 3),
 in which case the effect is still quite small.

In another set of plots, Fig.4, we fix $\rh$ and vary $\ri$ between
values  larger than $\rh$ to values much smaller than $\rh$.
   We see that the bulk of the left handed zero-mode  is ``pushed'' inside
 the small instanton core (well inside the Higgs core),
 while the right handed
zero-mode spreads over the instanton core, but its
 chracteristic radius is
still of the order $\ri$.  This behaviour is generic for the values of
Yukawa  coupling corresponding to masses of fermions in the standard
model up to $\sim 100 GeV$.

In conclusion, let us note that the
 non-spherical solutions (see eq. (8), (9))
can also be found numerically, but they would not  present any new
features  under the variations of the two scales discussed above.

\paragraph{Acknowledgements-}
 I wish to thank V. Kaplunovsky for suggesting
the study of zero-modes in instanton EW-type theories. Also, the useful
discussions with S. Gousheh and
S. Thomas  are gratefully acknowledged.

\newpage
\centerline{{\sc Figure Captions}}

\vspace{0.3in}

\paragraph{Fig. 1} - ($a$) Normalized left handed zero modes for \( \ri=0.00
02 GeV^{-1}\), and \(\rh=\ri,
 0.25 \ri\), \(10^{-1}\ri, 10^{-2}\ri\), superimposed. ($b$)
 Normalized right handed zero modes of fermions.
 The curve with lowest amplitude corresponds to
 $\ri=\rh$. The same general picture holds for left and right handed fermions
of
mass $0.01GeV\le m_F\le100GeV$.

\paragraph{Fig. 2} - Normalized left handed  ($a$) and right handed ($b$) zero
modes for \(
\ri=0.001 GeV^{-1}\), and \( \rh=\ri\),
\( 0.25 \ri\), \(10^{-1}\ri\), \(10^{-2}\ri\).

\paragraph{Fig. 3} - Normalized left handed ($a$) and right handed ($b$)
zero modes. $m_F=500GeV$. $\ri$ and $\rh$ are
the same as those given in Fig.2.

\paragraph{Fig. 4} - Normalized  left and right handed fermion zero-modes.
Here,
$\rh$ is kept fixed and ($a$) $\ri=2\rh$, ($b$) $\ri=\rh$, ($c$) $\ri=\rh/5$,
($d$)
$\ri=\rh/50$.
 In these plots, the coordinate $z$ is defined slightly
differently from that described in the text, namely, \(z=\tanh(r^2v^2)\), where
\(v=250GeV\).
 The vertical line represents the size of the instanton
in the $z$-coordinate.

\end{document}